\documentclass[a4paper,11pt]{article}
\usepackage{pos}
 
\usepackage{amsmath}

\usepackage{mathrsfs}
\usepackage{multirow}
\usepackage{multicol}
\usepackage{todonotes}
\usepackage{hyperref}
\usepackage{subcaption}

\usepackage{float}

\usepackage{cleveref}
\usepackage{xcolor,colortbl}
\usepackage{booktabs}

\title{Variance reduction with probing and Multilevel Monte Carlo in Lattice QCD}

\author[a]{Andreas Frommer}
\author*[a]{Jose Jimenez-Merchan}
\author[a]{Bruno Lang}
\author[a]{Mario Papace}
\author[b]{Gustavo Ramirez-Hidalgo}

\affiliation[a]{Bergische Universität Wuppertal\\
Fakultät für Mathematik und Naturwissenschaften\\
Gaußstraße 20, 42119 Wuppertal, Germany}
\affiliation[b]{J{\"u}lich Supercomputing Centre, \\Forschungszentrum J{\"u}lich GmbH, Wilhelm-Johnen-Straße 52428 Jülich, Germany}

\emailAdd{\{frommer,jimenezmerchan,lang,papace\}@uni-wuppertal.de}
\emailAdd{g.ramirez.hidalgo@fz-juelich.de}

\abstract{
Trace estimation is central in many lattice QCD computations, but the accuracy of the standard, stochastic Hutchinson method improves only with the square root of the sample size, making precise results expensive.

We investigate two complementary variance reduction strategies. First, \emph{multigrid multilevel Monte Carlo} uses a multigrid hierarchy to construct an unbiased multilevel estimator via recursive coarse grid corrections available from the multigrid hierarchy of the solver. Second, \emph{stochastic probing} uses distance-$d$ graph colorings; we propose a torus based coloring that requires substantially fewer colors than hierarchical probing at the same distance.

We test these approaches on two representative problems: the connected pseudoscalar correlator and disconnected fermion loops. For the connected pseudoscalar two-point function, the multilevel decomposition yields a variance reduction of up to $\mathcal{O}(10^5)$ at large time separations and translates into a clear cost reduction at fixed accuracy, thus confirming earlier results of \cite{Gruber-MG-LMA}. For the disconnected loops, in contrast, the multilevel decomposition provides only moderate gains, whereas probing combined with dilution delivers a substantial cost reduction that improves as the number of probing vectors is increased. Overall, the results highlight a pronounced complementarity: deflation schemes are most effective for observables dominated by long distance propagation, while probing is most effective for localized quantities.
}

\FullConference{
42nd International Symposium on Lattice Field Theory (Lattice 2025)\\
Tata Institute of Fundamental Research (TIFR), Mumbai, India\\
November 2-8 2025
}

\newcommand{\tr}{\mathrm{tr}}

\begin{document}
\maketitle

\section{Introduction} \label{sec:introduction}
We consider the problem of computing $\tr\left[B(t)D^{-1}(t, t') \right]$, where $D \in \mathbb{C}^{n \times n}$ is the discretized Wilson-Dirac operator in Lattice Quantum Chromodynamics (QCD). This problem arises when computing connected and disconnected contributions to hadronic correlation functions. The Wilson-Dirac matrix $D$ is a block-structured matrix with each lattice site represented by a $12\times12$ block corresponding to the internal degrees of freedom of spin ($4$ components) and color ($3$ components). $D^{-1}(t, t')$ is a block of the inverse Wilson-Dirac matrix correlating the time slices $t$ and $t'$, while $B(t)$ is an operator possibly acting on spin, color, and space indices at a fixed time slice $t$. The dimension of such operators is usually so large that explicitly computing the trace becomes impractical, which necessitates the use of stochastic methods.\\
Consider a stochastic vector $\eta \in \mathbb{C}^n$ with components $\eta_j$ that satisfy
\begin{equation}
    \label{eq:random-vectors-entries}
        \mathbb{E}[\eta_j] = 0, \qquad
        \mathbb{E}[\bar{\eta}_j \eta_k] = \delta_{jk}, \qquad \text{for} \quad i,k=1,...,n. \quad j\neq k.
\end{equation}
Then we have the following relation:
\begin{equation}
\label{eq:Hutchinson-identity}
    \tr(f(D)) = \mathbb{E}[\eta^{\dagger} f(D)\eta],
\end{equation}
where $f$ denotes any function that can be applied to $D$. Since we are interested in the inverse, we take $f$ to be the inverse. Using \cref{eq:Hutchinson-identity}, this yields the Hutchinson estimator \cite{Hutchinson1990} for $\tr(D^{-1})$:
\begin{equation}
    \label{eq:Hutchinson-estimator}
    \tr(D^{-1}) \approx \widehat{\tr}(D^{-1}) = \frac{1}{N} \sum_{i = 1}^N (\eta^{(i)})^\dag D^{-1}\eta^{(i)},
\end{equation}
\\
where $\eta^{(i)}$ are samples from $\eta$. Thus, in our computations we do not require the explicit construction of the full matrix $D^{-1}$, but we only need products of the form $D^{-1}\eta^{(i)}$, which can be obtained by solving a linear system with $D$ as the system matrix and $\eta^{(i)}$ as the right-hand side. \\
A common choice for the random vectors $\eta^{(i)}$ in \cref{eq:Hutchinson-estimator} is to take them as \emph{Rademacher vectors}, meaning that each component is independently equal to $+1$ or $-1$ with probability $\frac{1}{2}$ each. 
In this case, the variance of the Hutchinson estimator is given by \cite{Bernardson1994}:
\begin{equation}
\label{eq:hutchinson-variance}
    \mathbb{V}[\eta^{\dagger} D^{-1} \eta] = \frac{1}{2}\|\text{offdiag}(D^{-1} + (D^{-1})^T)\|_F^2
\end{equation}
where offdiag$(A)$ is the off-diagonal part of the matrix $A$ and $A^T$ denotes the (non-conjugate) transpose. As a Monte-Carlo sampling method, the accuracy of the estimator \cref{eq:Hutchinson-estimator} improves only as $\frac{1}{\sqrt{N}}$, making the method too expensive when higher precision is required. In \cref{sec:MGMLMC} we describe a multilevel strategy that addresses this limitation. In the same spirit, we in \cref{sec:probing}  introduce a variance reduction method based on graph coloring with a new coloring scheme. Finally, in \cref{sec:results} we present numerical results for the  connected pseudoscalar correlator observables and disconnected fermion loops. The results show that observables influenced by low modes benefit most from deflation, whereas graph coloring is better for local, short-range fluctuations.

\section{Variance reduction: multigrid multilevel Monte Carlo}
\label{sec:MGMLMC}

Since the Frobenius norm and the singular values of any matrix $A\in\mathbb{C}^{n\times n}$ are related by $\|A\|_F^2 = \sum_{i=1}^n \sigma_i^2$,\, the variance in \cref{eq:hutchinson-variance} is dominated by low-mode (near-null-kernel) contributions. A possible remedy is \emph{deflation}: one selects a subspace spanned by (exact or approximate) low eigenmodes and projects it out \cite{Gambhir2017}. This can be done, for example, using the symmetrized Dirac operator, $Q=\Gamma_5D$. Let $V_k=[v_1|\cdots|v_k]$ contain $k$ low-mode approximations of $Q$. With the orthogonal projector $\Pi = V_kV_k^\dagger$ we can split

\begin{equation}
\label{eq:defl-split}
\tr(D^{-1}) = \tr\bigl((I-\Pi)D^{-1}\bigr) + \tr\!\bigl(\Pi D^{-1}\bigr),
\end{equation}
\\
so that random sampling is applied to the deflated component with a reduced number of samples, while the low-mode part is relatively inexpensive to compute, usually by exact computation involving $k$ linear solves. 
The multigrid multilevel Monte Carlo (multigrid MLMC) method \cite{FrommerMultilevel} can be viewed as a recursive, level-wise generalization of this idea: instead of deflating with a single global subspace, we deflate successively using coarse-grid corrections provided by a multigrid hierarchy. It is based on multilevel Monte Carlo (MLMC) \cite{Giles2015}, a general framework relying on an additive decomposition of a stochastic variable $X$ into $L$ stochastic variables $X_l$, i.e, $X=\sum_{l=1}^L X_l$ such that the cost  $C_l$ to compute a sample from $X_l$ is small when the variance $V_l$ of $X_l$ is large. The cost-minimizing choice for the number of samples $N_l$ for $X_l$ to reach a target variance $\epsilon^2$ is 
\begin{equation}
\label{eq:MLMC-Nl}
N_l = \frac{1}{\epsilon^2}\sqrt{\frac{V_l}{C_l}}\sum_{i=1}^L \sqrt{V_iC_i}.
\end{equation}
\\
Examples of this general framework include the {\em frequency splitting} approach of \cite{Giusti2019}, \cite{Whyte202410892} and the {\em truncated solver method} \cite{Bali2010}.  In a multigrid setting we use the operator hierarchy $D_1=D$, $D_l\in\mathbb{C}^{n_l\times n_l}$, for $l=1,...,L$, together with prolongators $P_l$ and restrictors $R_l$ between levels $l$ and $l+1$. For aggregation-based multigrid solvers in lattice QCD (e.g.,\ DD$\alpha$AMG \cite{DDalphaAMG-paper}) one has $R_l=P_l^\dagger$ with $P_l^\dagger P_l = I$ and a Galerkin coarse operator

\begin{equation}
\label{eq:galerkin}
D_{l+1} = P_l^\dagger D_l P_l,\qquad l=1,\ldots,L-1.
\end{equation}
\\
Multigrid MLMC realizes a recursive deflation using the oblique projector
\begin{equation}
\label{eq:oblique-projector}
\Pi_l = P_l D_{l+1}^{-1} P_l^\dagger D_l,
\end{equation}
\\
which induces the level splitting
\begin{equation}
\label{eq:level-split}
D_l^{-1} =
\Bigl(D_l^{-1} - P_l D_{l+1}^{-1} P_l^\dagger\Bigr)
+ P_l D_{l+1}^{-1} P_l^\dagger .
\end{equation}
\\
Denoting
\begin{equation}
\label{eq:mlmc-diff-op}
M_l := D_l^{-1} - P_l D_{l+1}^{-1} P_l^\dagger = (I-\Pi_l)D_l^{-1},
\end{equation}
\\
and applying the splitting recursively, yields 

\begin{equation}
\label{eq:oblique_multigrid-mlmc_short}
\tr(D^{-1}) =
\sum_{l=1}^{L-1}\tr(M_l) + \tr(D_L^{-1}).
\end{equation}
\\
Each term can now be estimated stochastically (optionally combined with additional variance reduction methods), giving the
unbiased multigrid MLMC estimator

\begin{equation}
\label{eq:multigrid MLMC-est}
\widehat{\tr}(D^{-1})
= \sum_{l=1}^{L-1}\widehat{\tr}(M_l) + \widehat{\tr}(D_L^{-1}).
\end{equation}
\\
The rationale is that the prolongators $P_l$ are built from test vectors approximating low modes of $D$, and due to local coherence \cite{Luscher:Local-Coherence} they capture global near-null-kernel information. As a consequence, the variance at the expensive fine levels is reduced and shifted to coarser levels, where solves are significantly cheaper. In practice, the transfer operators $P_l$ and $R_l$ are constructed during a \emph{setup phase} from $N_{\mathrm{tv}}$ \emph{test vectors} $v_1,\ldots,v_{N_{\mathrm{tv}}} \in \mathbb{C}^{n_l}$ that approximate eigenvectors associated with the smallest eigenvalues of $D_l$.

A decomposition similar to \cref{eq:level-split} has recently been used within Wick contractions of hadronic two-point correlators in \cite{Gruber-MG-LMA}, reporting speed-ups up to $30\times$ compared to plain Hutchinson. 
\section{Probing} \label{sec:probing}
An alternative approach used to reduce the variance of the stochastic estimation of traces is \emph{probing} \cite{Wilcox-probing}, a technique that takes advantage of the sparsity pattern of $D$ to reduce the variance in \cref{eq:hutchinson-variance} by annihilating parts of the off-diagonal of $D^{-1}$. Probing is a deterministic method that exploits the distance-$d$ coloring of the graph associated with $D$ (where all nodes within a distance $d$ in the graph must have different colors), to construct structurally orthogonal probing vectors to extract specific non-zero entries of the matrix. 

Probing can be combined with stochastic trace estimation to develop {\em stochastic probing}, an improved trace estimation technique whose key idea is to replace purely random vectors \cref{eq:random-vectors-entries} with random probing vectors coming from a distance-$d$ coloring of the matrix graph and that respect the matrix sparsity pattern.  Assuming that we have computed a distance-$d$ coloring of the graph of $D$, which results in $n_c$ colors, probing vectors are built in the following way:
\begin{equation}
    v_{j}(i) = \begin{cases} \pm1, & \text{if color}(i) = j   \\ 0, & \text{otherwise} \end{cases}.
    \label{probing vectors}
\end{equation}
\\
We choose the nonzero entries of the probing vectors to be drawn from the same probability distribution of Rademacher vectors, so that they satisfy the conditions in \cref{eq:random-vectors-entries}. In this way, probing vectors can replace Rademacher vectors in \cref{eq:Hutchinson-estimator}, which becomes:
\begin{equation}
    \label{eq:Hutchinson-probing-estimator}
    \text{tr}(D^{-1}) \approx \sum_{j = 1}^{n_c}\frac{1}{N}\sum_{i = 1}^N v^{(i)}_jD^{-1}v^{(i)}_j
\end{equation}
where the trace associated to every color is independently estimated. We refer to \cite{FroRinSch2023}, e.g., for an analysis of stochastic probing in connection with a quantification of the decay of the entries of $D^{-1}$away from the graph of $D$.

Computing a distance-$d$ coloring of a graph can be very costly because, unless the graph exhibits particular structural properties that can be exploited, one must resort to a greedy approach in which, for each node in the graph, all other nodes are examined one by one to determine whether they are neighbors. Hierarchical probing \cite{Stathopoulos-hp} provides a \emph{nested} sequence of probing vectors that progressively cancels contributions from matrix entries at increasing graph distances. In practice it targets distances that grow in powers of two and is especially efficient on regular lattices, where the construction can be done without explicitly computing an expensive greedy distance-$d$ coloring. Because the vector sets are nested, samples generated at a given distance can be extended (rather than discarded) when moving to a larger distance. In the following, we propose a different coloring scheme that is not restricted to distances that are powers of two and uses significantly fewer colors than those produced by hierarchical probing.

Given our focus on physical observables in lattice QCD, we can take advantage of the structure of the Dirac matrix. When broken down into its 12 × 12 block form, this matrix corresponds to the adjacency matrix of a toroidal lattice with nearest-neighbor interactions. By exploiting the  connectivity structure of such a torus, we can obtain a distance-$d$ coloring by:

\begin{equation}
    \label{eq:coloring-formula}
    c(i) = \sum_{j=1}^{n_D} \sigma_jx_j \hspace{0.2cm} \text{mod} \hspace{0.2 cm}n_c
\end{equation}
where $c(i)$ is the color assigned to the lattice site $i$ with coordinates ($x_1, x_2, ..., x_{n_D}$), $n_D$ is the spacetime dimension (which is 3 on a time slice and 4 on the whole lattice), $n_c$ is the number of colors and ($\sigma_1, \sigma_2, ..., \sigma_{n_D}$) are coefficients that depend on the dimensions of 
the lattice and on the coloring distance. An example for the $64 \times 32^3$ torus is given in Table~\ref{tab:bruno_vs_hp_nc}.
The coefficients $\sigma_i$ which lead to a small number of colors $n_c$ must be determined computationally via an appropriate exhaustive search 
technique. Note that they can be determined and tabled once and then kept for ever for given lattice dimensions.
\begin{table}
\centering
\

\begin{minipage}{0.48\linewidth}
\centering
\begin{tabular}{|c|c|c|}
\hline
\multicolumn{3}{|c|}{$n_D=3$}\\ \hline
$d$ & $n_c$ & $n_c^{(HP)}$\\ \hline
1 & 2 & 2\\ \hline
2 & 8 & 16\\ \hline
3 & 16 & -\\ \hline
4 & 32 & 128\\ \hline
\end{tabular}
\end{minipage}
\hfill
\begin{minipage}{0.48\linewidth}
\centering
\begin{tabular}{|c|c|c|}
\hline
\multicolumn{3}{|c|}{$n_D=4$}\\ \hline
$d$ & $n_c$ & $n_c^{(HP)}$\\ \hline
1 & 2 & 2\\ \hline
2 & 10 & 32\\ \hline
3 & 16 & -\\ \hline
4 & 64 & 512\\ \hline
\end{tabular}
\end{minipage}
\caption{Number of colors associated with different coloring distances obtained from \cref{eq:coloring-formula} and from hierarchical probing for a $64 \times 32^3$ torus. Here $d$ denotes the coloring distance, $n_c$ the number of colors obtained from \cref{eq:coloring-formula}, while $n_c^{(HP)}$ is the number of coloring in hierarchical probing. \label{tab:bruno_vs_hp_nc} }
\end{table}

The variance of the trace estimation can be further decreased by incorporating dilution \cite{Foley2005} in addition to probing, which refers to the process of selectively decoupling the internal degrees of freedom when constructing the stochastic sources used in the estimation. By doing so, one effectively eliminates unwanted cross-talk between components. Consequently, the number of independent estimates in \cref{eq:Hutchinson-probing-estimator} is increased by a factor determined by the selected dilution scheme (12 for full dilution), because the estimate for each color is additionally decomposed into its spin-color components. The next section presents the results of the application of probing and its combinantion with dilution to the estimation of physical observables.  

\section{Results on physical observables} \label{sec:results}

This section presents numerical results for connected and disconnected contributions in correlation functions and uses them to illustrate that multigrid MLMC and probing achieve variance reduction through different mechanisms. Multigrid MLMC is most effective on observables which are strongly influenced by low modes, whereas probing is most effective when the dominant fluctuations are short ranged and local. 

We performed all experiments using the configuration from \cref{tab:configuration}. The multigrid setup
uses three levels with $N_{tv}=28$ test vectors for the construction of the transfer operators. A fixed sample size $N=500$ for the estimation for all variance estimates.

\begin{table}
\centering
\caption{Lattice size at all levels used in the multigrid hierarchy for a configuration from the IV ensemble generated by the Regensburg QCD (RQCD) collaboration \cite{Regensburgcollab}. }
\label{tab:configuration}
\begin{tabular}{lccc|cc}
\toprule
Ensemble & $T \times L^3$ & Level 2 & Level 3  & $m_\pi$ [MeV] & $a$ [fm] \\
\midrule
IV   & $64 \times 32^3$  & $16 \times 8^3$  & $8 \times 4^3$             & $295$ & 0.071 \\
\bottomrule
\end{tabular}
\end{table}

\subsection{Connected case: pseudoscalar two-point function}

As a representative connected contribution we consider the pseudoscalar two-point function
\begin{equation}
\label{eq:connected-results-G}
G(t)=\frac{1}{T}\sum_{t'} \mathrm{tr}\!\left[D^{-1}(t+t',t')\,\Gamma_5\,D^{-1}(t',t+t')\,\Gamma_5\right],
\end{equation}
where the trace is over space, spin and color.

Similarly to \cite{Gruber-MG-LMA}, we use the multigrid decomposition of \cref{eq:level-split} and denote by $D_l^{-1}(t+t',t')$ the $(t+t',t')$ time slice block of $D_l^{-1}$ 
and define level operators

\begin{equation}
\label{eq:connected-results-Bl}
B_l(t+t',t')=\hat P_l\!\left(D_l^{-1}(t+t',t')-P_l D_{l+1}^{-1}(t+t',t')P_l^\dagger\right)\!\hat P_l^\dagger,\quad l=1,\dots,L-1,
\end{equation}
\\
and
\begin{equation}
\label{eq:connected-results-BL}
B_L(t+t',t')=\hat P_L D_L^{-1}(t+t',t')\hat P_L^\dagger,
\end{equation}
where $\hat{P}_l = P_1\cdots P_l$ and $\hat{R}_l = R_l\cdots R_1$ are the accumulated transfer operators. With these definitions, the correlator (\ref{eq:connected-results-G}) admits the multilevel representation
\begin{equation}
\label{eq:connected-results-Gsum}
G(t)=\sum_{i,j=1}^{L} G_{i,j}(t),
\end{equation}
where
\begin{equation}
\label{eq:connected-results-Gij}
G_{i,j}(t)=\frac{1}{T}\sum_{t'=1}^{T}\mathrm{tr}\!\left[B_i(t+t',t')\Gamma_5\,B_j(t',t+t')\Gamma_5\right] \qquad \text{for} \quad i,j=1,...,L.
\end{equation}
The $\Gamma_5$ insertions prevent a cyclic reordering of factors that would change the relative positions of the $B_i$ and $B_j$ factors; therefore the order of the accumulated transfer operators must be preserved.

\begin{figure}[t]
\centering
\includegraphics[scale=0.32]{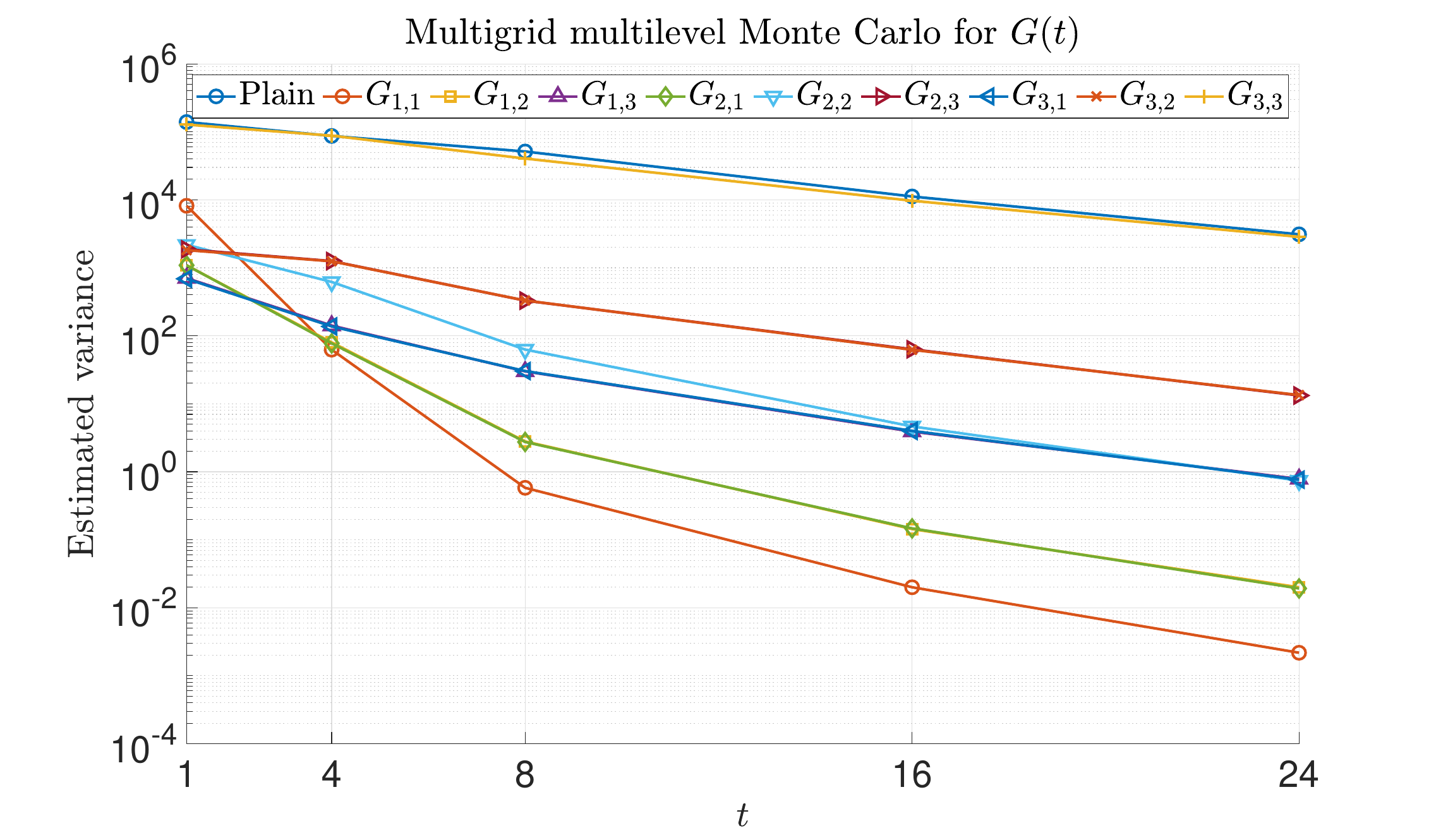}
\caption{\label{fig:connected-traditional} Comparison between the plain Hutchinson estimator and the multigrid Multilevel Monte Carlo estimator based on oblique projections.}
\end{figure}

\Cref{fig:connected-traditional} compares the variance of the plain Hutchinson estimator for $G(t)$ with the variances of the components $G_{i,j}$ in the oblique multilevel decomposition \cref{eq:connected-results-Gsum}. The dominant fine-level contribution $G_{1,1}$ exhibits a pronounced variance reduction relative to plain Hutchinson: already at $t=1$ the reduction is $\mathcal{O}(10^3)$ and it increases with the time separation $t$, reaching about $\mathcal{O}(10^5)$ at $t=24$. In contrast, the coarsest and cheapest term $G_{3,3}$ carries the largest variance among the $G_{i,j}$ contributions across all $t$. This allocation is well suited for multilevel Monte Carlo, since the residual variance is concentrated in inexpensive coarse level terms that can be sampled at low cost, while the expensive fine level terms have strongly reduced variance.

\subsection{Fermion loop disconnected contributions}

\begin{figure}[t]
\centering
\includegraphics[width=0.5\textwidth]{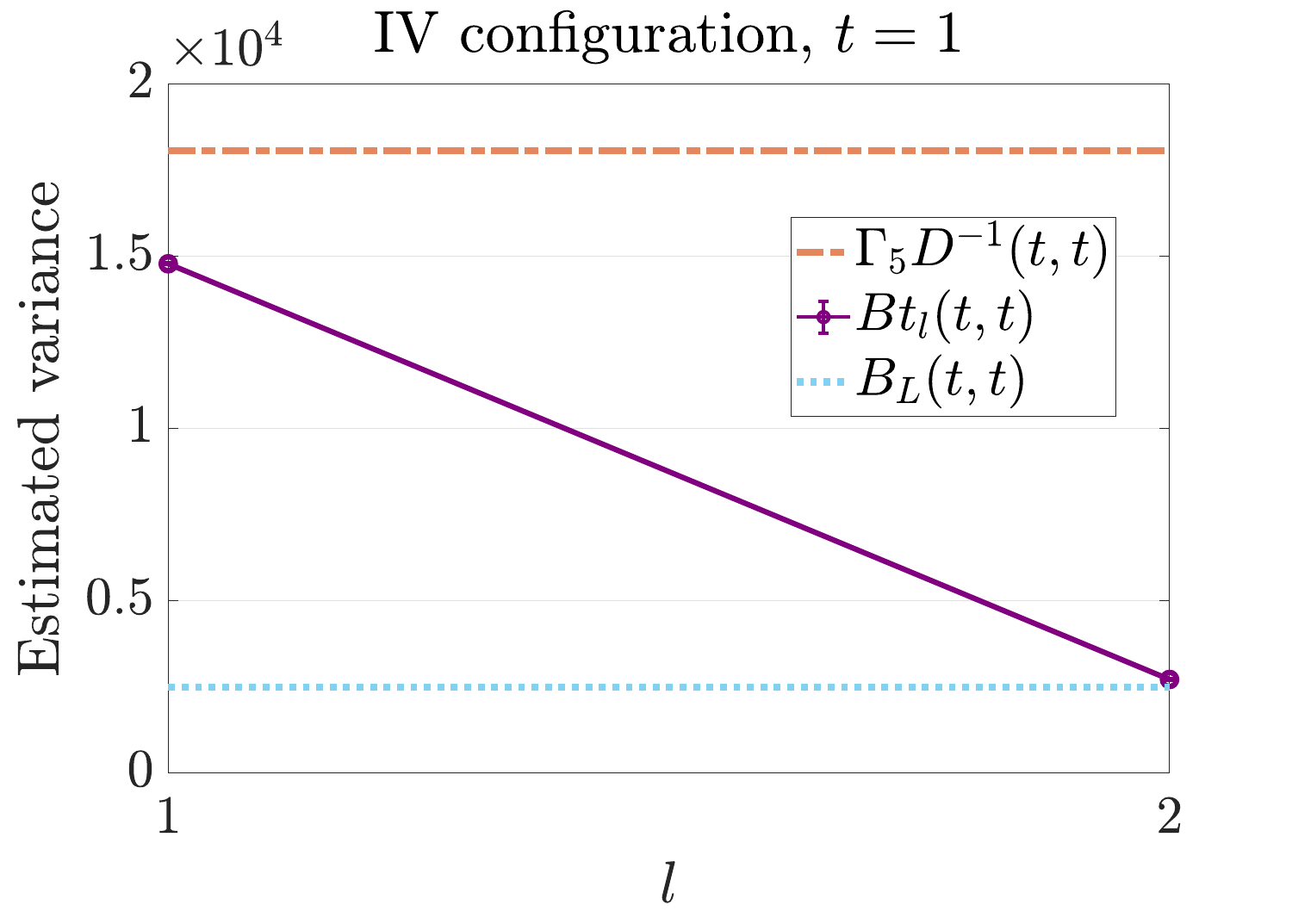}
\caption{Estimated variances for $\mathrm{tr}(\Gamma_5(t)D^{-1}(t,t))$ on the IV configuration at $t=1$, comparing plain Hutchinson to the multilevel terms.}
\label{fig:disconnected-mlmc-iv}
\end{figure}

For disconnected contributions we consider the trace
\begin{equation}
\label{eq:disconnected-results-O}
O(t)=\mathrm{tr}\!\left[\Gamma_5(t)\,D^{-1}(t,t)\right],
\end{equation}
which enters disconnected correlation functions such as
\begin{equation}
\label{eq:disconnected-results-C}
C(\tau)=\left\langle \frac{1}{T}\sum_{t=0}^{T-1} O(t)\,O(t+\tau)\right\rangle_g .
\end{equation}
\\
To construct a multilevel hierarchy for this quantity, we use the decomposition
\begin{equation}
\label{eq:disconnected-results-trace-decomp}
\mathrm{tr}\!\left(\Gamma_5(t)\,D^{-1}(t,t)\right)=\sum_{l=1}^{L}\mathrm{tr}\!\left(\Gamma_5\,B_l(t,t)\right),
\end{equation}
where $B_l(t,t)$ is defined analogously to \cref{eq:connected-results-Bl,eq:connected-results-BL}, now restricted to the $(t,t)$ time-slice block. The measurements were done for $t=1$. The finest level term shows a moderate reduction in variance relative to plain Hutchinson, while a substantial contribution remains in the coarser terms. In the current setup, the reduction at the finest level is not sufficient to compensate for the additional overhead of the multilevel estimator, and we therefore do not obtain a net cost improvement over plain Hutchinson for this disconnected loop contribution. The corresponding variances are shown in \cref{fig:disconnected-mlmc-iv}.

\begin{figure}[t]
\centering
\includegraphics[width=0.78\textwidth]{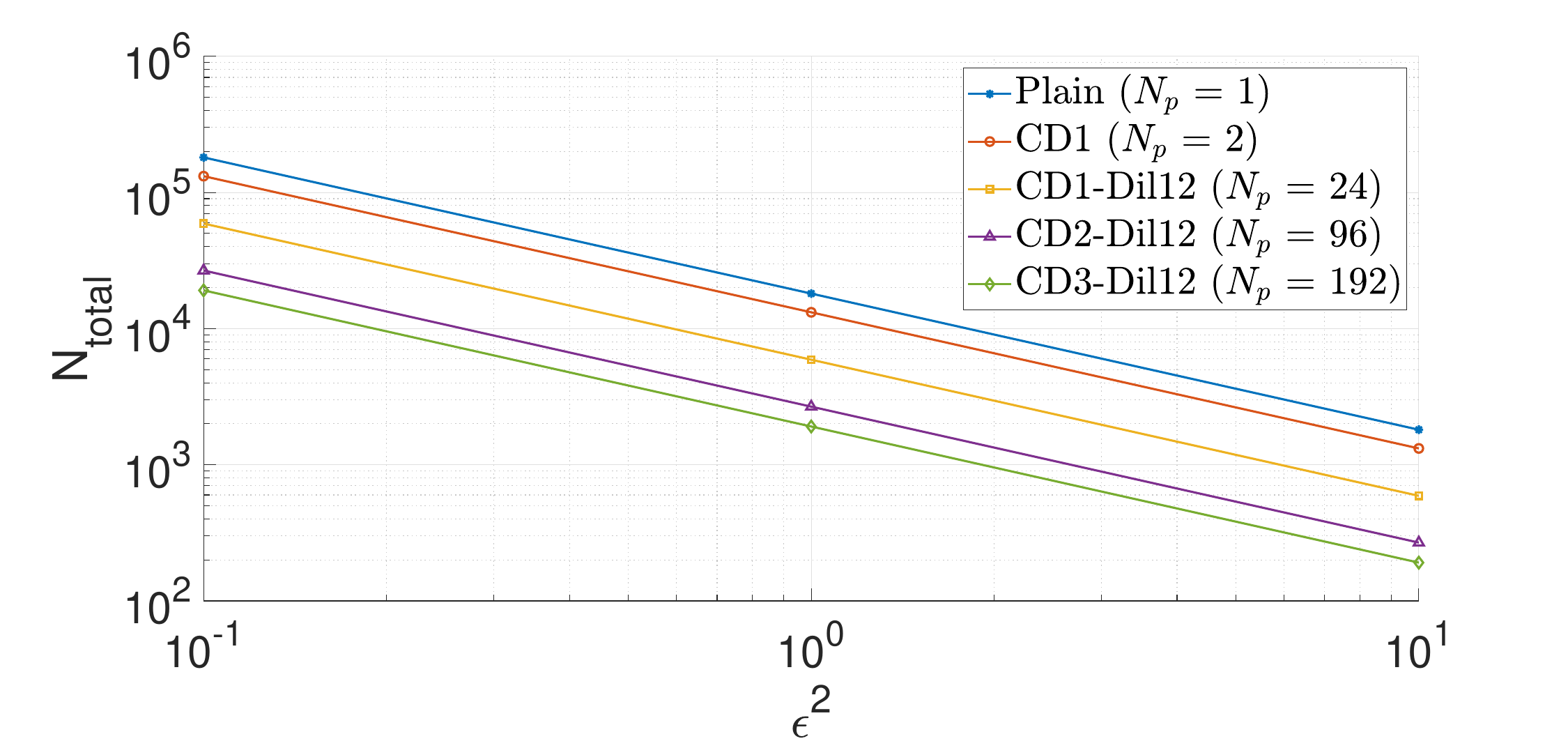}
\caption{Total number of linear system solves ($\text{N}_{\text{total}}$) required to reach target variance $\varepsilon^2$ for the disconnected correlator. Coloring distances ($\textbf{CD}$) from 1 to 3 are explored in combination with full dilution ($\textbf{Dil12}$) where all the 12 spin and color degrees of freedom are decoupled.}
\label{fig:disconnected-probing}
\end{figure}

We next consider probing for the disconnected loop contribution, combined with dilution. The probing vectors are generated using the coloring in \cref{eq:coloring-formula}, which requires a factor of four fewer colors than hierarchical probing for the same coloring distance. \Cref{fig:disconnected-probing} shows a substantial reduction in the total number of solves needed to reach a prescribed target variance when compared with plain Hutchinson, and the improvement grows as the number of probing vectors $N_p$ increases.

\section{Conclusions and outlook} \label{sec:conclusions}

Taken together, our numerical results show that multigrid MLMC and probing address complementary sources of stochastic variance. For the connected correlator, multigrid MLMC yields a clear reduction in the computational cost at fixed accuracy, consistent with the fact that long distance propagation is strongly influenced by components that are efficiently handled by the multilevel hierarchy. For the disconnected loop, in contrast, multigrid MLMC provides only a moderate variance reduction on the finest level and does not yet outperform plain Hutchinson in overall cost, whereas probing combined with dilution achieves a substantial reduction in cost. This indicates that the practical effectiveness of the two approaches is closely tied to the structure of the observable: long-distance correlators benefit most from deflation schemes, while local in time quantities benefit most from variance reduction methods targeted to locality.

Quantitatively, both approaches can deliver cost reductions of about one order of magnitude in the regimes where they are best matched to the observable, corresponding to variance reductions on the order of $10^{4}$ at fixed sample size.

Several directions remain to be explored. On the probing side, it is important to benchmark the three dimensional torus coloring from \cref{eq:coloring-formula} against hierarchical probing in a controlled comparison at fixed cost. On the multigrid MLMC side, further improvements may be possible by incorporating additional exact low modes or by using more accurate approximations to the low-eigenmode subspace in the transfer operator construction. Finally, the observed complementarity motivates studying combined strategies, in particular the integration of multigrid MLMC with the proposed probing approach, with the goal of simultaneously targeting long-distance and short-distance contributions to the variance.

\section*{Acknowledgments}
\label{sec:acknowledgements}
This work is supported by the German Research Foundation (DFG) research unit FOR5269 ”Future methods
for studying conﬁned gluons in QCD”. The computations for the measurements of the variance were carried out on the PLEIADES cluster at the University of Wuppertal, which was supported by the Deutsche Forschungsgemeinschaft (DFG, grant No. INST 218/78-1 FUGG) and the Bundesministerium für Bildung und Forschung (BMBF).
G.R-H. acknowledges financial support from the EoCoE-III project, which has received funding from the European High Performance Computing Joint Undertaking under grant agreement No. 101144014.
M.P. acknowledges AQTIVATE (\url{https://aqtivate.ucy.ac.cy/}) which has received funding from the European Union’s research and innovation programme under the Marie Skłodowska-Curie Doctoral Networks action and Grant Agreement No 101072344.
The authors gratefully acknowledge the Gauss Centre for Supercomputing e.V. (\url{www.gauss-centre.eu}) for funding this project by providing computing time through the John von Neumann Institute for Computing (NIC) on the GCS JUWELS at Jülich Supercomputing  Centre (JSC).

\bibliographystyle{JHEP.bst}
\bibliography{refs}

\end{document}